\documentclass[12pt]{article} 

\ifx\ltimes\undefined
\newcommand{\ltimes}{{\kern3pt\hbox{\vrule width 0.4pt height 5.30pt
depth .0pt}\kern-1.76pt\times\kern1pt}} \fi

\ifx\lrtimes\undefined
\newcommand{\rtimes}{{\kern1pt\times\kern-4.76pt\kern3pt\hbox{\vrule width 0.4pt height 5.30pt
depth .0pt}}} \fi

\def\D{{\cal D}}

\def\G{{\cal G}}

\def\cL{{\mathcal{L}}}
\def\M{{\cal M}}
\def\N{{\cal N}}

\def\d{\partial}
\def\+{{+\!\!\!+}}
\def\pp{\mbox{\tiny${}_{\stackrel\+ =}$}}

\def\wt{\widetilde}
\def\half{\frac{1}{2}}

\def\diag{\textstyle{\rm{diag}}}

\def\span{\textstyle{\rm{span}}}

\def\Qp{{\bf Q}_+}
\def\Qm{{\bf Q}_-}

\def\bid{\hbox{1\hspace{-0.04in}I}} %blackboard bold 1
\def\bR{\hbox{I\hspace{-0.04in}R}} %blackboard bold R

\newcommand{\nc}{\newcommand}
\nc{\beq}{\begin{equation}}
\nc{\eeq}[1]{\label{#1}\end{equation}}
\nc{\ber}{\begin{eqnarray}}
\nc{\eer}[1]{\label{#1}\end{eqnarray}}

\newcommand{\Section}[1]{\section{#1} \setcounter{equation}{0}}

\begin{document}

\begin{center}
                       
                                \hfill   hep-th/0606024\\
                                \hfill   YITP-06-25\\
                                \hfill   Imperial/TP/06/RAR/03\\
                                \hfill   QMUL-PH-06-07\\

\vskip .3in \noindent

\vskip .1in

{\large \bf{Worldsheet boundary conditions in Poisson-Lie T-duality}}
\vskip .2in

{\bf Cecilia Albertsson}$^{a,}$\footnote{e-mail
 address: cecilia@yukawa.kyoto-u.ac.jp}
 and  {\bf Ronald A.\ Reid-Edwards}$^{b,}$\footnote{e-mail
 address: r.reid-edwards@imperial.ac.uk} \\

\vskip .15in

\vskip .15in
$^a${\em Yukawa Institute for Theoretical Physics, Kyoto University, \\
Kyoto 606-8502, Japan}
\vskip .15in
$^b${\em The Institute for Mathematical Sciences, Imperial College London,\\
53 Prince's Gate, London SW7 2PG, UK\\
and\\
Department of Physics, Queen Mary, University of London,\\
Mile End Road, London E1 4NS, UK}\\

\bigskip

\vskip .1in

\end{center}
\vskip .4in

\begin{center} {\bf Abstract } 
\end{center}
\begin{quotation}\noindent 
We apply canonical Poisson-Lie T-duality transformations
to bosonic open string worldsheet boundary conditions, showing
that the form of these conditions is invariant at the classical level, and therefore they are compatible with Poisson-Lie T-duality. In particular the conditions for conformal invariance are automatically preserved, rendering also the dual model conformal. The boundary conditions are defined in terms of a gluing matrix which encodes
the properties of D-branes, and we derive the duality map for this matrix. 
We demonstrate explicitly the implications of this map for D-branes in
two non-Abelian Drinfel'd doubles.

\end{quotation}
\vfill
\eject

\tableofcontents

\Section{Introduction}

T-duality in string theory may be realised as a transformation acting on the two-dimensional nonlinear sigma model \cite{Ooguri,BorlafLozano,Hori,HKLR86,Rocek,Hassan2,Alvarez,Dorn,Forste}. This model describes the worldsheet theory of a string propagating on some target manifold $\M$ equipped with a riemannian metric $g_{\mu\nu}$, an antisymmetric B-field $B_{\mu\nu}$ and a torsion $H=dB$. Originally the condition for this realisation to be possible was that $\M$ have some isometry group $G$ (i.e., a group action $G$ on $\M$ under which the target space metric is invariant) which leaves the sigma model invariant. The dual model is then obtained by gauging the isometry to obtain a first-order parent action and integrating out gauge fields. The requirement that the background be isometric is a rather severe restriction, making it a challenging problem to prove T-duality for models where no isometry exists. Moreover, for non-Abelian isometry groups, the application of this technique is not symmetric in the sense that one does not necessarily recover the original theory by repeating the procedure \cite{HKLR86,DeWit,Giveon}.

Klim\v{c}\'{\i}k and \v{S}evera \cite{KS1} proposed a generalisation of T-duality to what has come to be known as Poisson-Lie T-duality, which allows the duality to be performed on a target space without isometries. Instead the background satisfies the Poisson-Lie condition, which is a restriction on the Lie derivative $\cL_{v^a} E_{\mu\nu}$ of the background tensor $E_{\mu\nu}\equiv g_{\mu\nu}+B_{\mu\nu}$ with respect to $G$-invariant vector fields $v^a$ on $\M$, replacing the isometry condition $\cL_{v^a} E_{\mu\nu}=0$. The Poisson-Lie condition is necessary for the existence of a well-defined dual worldsheet if the two dual target spaces are both Poisson-Lie group manifolds whose Lie algebras constitute a \emph{Drinfel'd double}.

A Poisson-Lie group is a Lie group with a Poisson structure that is compatible with the group operation. Every Poisson-Lie group is stratified into symplectic leaves with symplectic forms induced by the natural Poisson bracket on the group \cite{STS,Malkin,Chari}.
A Drinfel'd double \cite{Drinfeld2,Drinfeld1} is a Lie algebra $\D$ which decomposes into the direct sum, as vector spaces, of two maximally isotropic  Lie subalgebras $\G$ and $\wt \G$, each corresponding to a Poisson-Lie group ($G$ and $\wt G$), such that the subalgebras are duals of each other in the usual sense, i.e., $\wt \G=\G^*$. There is a complete classification of all real four- and six-dimensional Drinfel'd doubles \cite{Hlavaty1,Hlavaty2,Hlavaty3}, but not of the eight-dimensional ones.
If the target manifold $D/\wt G \cong G$ and its dual $G\backslash D \cong \wt G$ define a Drinfel'd double, then the Poisson-Lie condition translates into a flat-curvature condition, schematically $dJ+J\wedge J=0$,
on the Noether current $J$ generating the left-action of $G$ on itself in the sigma model. This is the condition for the worldsheet to be horizontally liftable into $D$ and hence to have a well-defined dual in $\wt G$. The lift defines a dressing action of $G$ on $\wt G$ and the ends of the open string on the dual target turn out to be confined to the orbits of the dressing action, which coincide with symplectic leaves \cite{KS2}. For example, in the special case of a WZW model they are (twisted) conjugacy classes \cite{Stanciu,Klimcik2}.

When performing traditional T-duality in the presence of an isometry, one finds equations of motion for the fields in the parent action, which constitute a map from the fields in the original model to those of the dual one. These are the canonical transformations, which if known can be applied directly to the fields in the model, without going to the trouble of gauging etc. In particular, they can be used to find the duals of the worldsheet boundary conditions for the open string. Such transformations exist also for Poisson-Lie T-duality \cite{Sfetsos1}, and they can be obtained as field equations of a parent action on the double as shown in \cite{vonUnge,KS4}. In this paper we apply the Poisson-Lie canonical transformations to the boundary conditions in order to find their Poisson-Lie T-duals.
 
The structure of the paper is as follows. In section \ref{prelim} we review the
open string worldsheet boundary conditions of the nonlinear sigma model
as derived in \cite{ALZ1,ALZ2,ALZ3}. These conditions are required for worldsheet $\N$=1 superconformal symmetry, but they were shown in \cite{ALZ4} to be necessary also for bosonic Abelian T-duality. Our analysis includes the full set of conditions even though the model is bosonic, partly with a view to future study of the supersymmetric theory, and partly because Abelian T-duality arises as a special case. We furthermore summarise those aspects of
Poisson-Lie T-duality relevant to our analysis, in particular recalling the (canonical) transformations as given by Klim\v{c}\'{\i}k and \v{S}evera \cite{KS1} and by Sfetsos \cite{Sfetsos1,Sfetsos2}.
In section~\ref{bconditions} we apply these
transformations to the worldsheet boundary conditions, deriving the duality map of the gluing matrix $R$ which defines the relation between left- and right-movers on the worldsheet boundary. We show that the form of the boundary conditions is invariant under this map, and in particular that conformal invariance is satisfied also on the dual side. In section~\ref{examples} we work out several examples explicitly: $U(1)^n$ Abelian T-duality; the semi-Abelian double (i.e., traditional non-Abelian T-duality); a simple non-Abelian double, namely the Borel algebra $\D=gl(2,\bR)$; and the three-dimensional double of Sfetsos \cite{Sfetsos3}. We conclude in section~\ref{conclusions} with a summary and outlook.

\Section{Preliminaries}
\label{prelim}

\subsection{Worldsheet boundary conditions}

Consider the bosonic nonlinear sigma model describing open strings
propagating on a Poisson-Lie group manifold $G$ with a general background tensor
$E_{\mu\nu} \equiv g_{\mu\nu} + B_{\mu\nu}$ (where $g_{\mu\nu}$ is a
metric on $G$ and $B_{\mu\nu}$ an antisymmetric
B-field):
\beq
S \equiv  \int d^2\xi \,\, \d_{\+} X^\mu \d_{=}X^\nu E_{\mu\nu}(X)\,.
\eeq{defSX}
Here $\d_{\pp} X^\mu$ denote derivatives, with respect to the worldsheet
coordinates $\xi^{\pm}$, of target space coordinates $X^\mu$.
In general the target space is locally a product $G\times \M_0$, where the fields
depending only on elements of $\M_0$ and not taking part in the duality transformation
are called spectators. Here we ignore spectator fields, but the inclusion
of them in the analysis is straightforward.

The worldsheet boundary is by definition confined to a D-brane. Its properties are encoded in the relation on the boundary between left- and right-moving worldsheet fields, which may be expressed as
\beq
\d_{=}X^\mu = R^\mu_{\,\,\,\nu} \d_{\+} X^\nu \,,
\eeq{origbc0}
for some gluing matrix $R^\mu_{\,\,\,\nu}$, satisfying
a set of conditions which we now discuss. In the $\N$=1 supersymmetric model the corresponding conditions state that, for the model to be consistent and for superconformal symmetry to be preserved on the boundary, the brane must be a well-defined smooth submanifold of the target manifold supporting a two-form defined by the B-field \cite{ALZ1,ALZ2}. The properties of the two-form determine whether the brane is Lagrangian, symplectic, or a more general type of submanifold.
Although the conditions for consistency of the bosonic model are less stringent in general, it turns out that one can gauge an Abelian isometry in the bosonic model only if (the bosonic versions of) the $\N$=1 conditions are satisfied along the directions of the isometry \cite{ALZ4}. With this in mind, and to allow a future straightforward extension to the supersymmetric model, we take into account all the boundary conditions derived in \cite{ALZ1,ALZ2}. They read as follows.

First, conformal symmetry requires $R^\mu_{\,\,\,\nu}$ to preserve the metric,
\beq
R^\rho_{\,\,\,\mu} g_{\rho\sigma} R^\sigma_{\,\,\,\nu} =g_{\mu\nu} \,.
\eeq{RgR0}
Next, vectors normal to the brane, which we refer to as Dirichlet vectors,
are eigenvectors of $R^\mu_{\,\,\,\nu}$ with eigenvalue $-1$.
In terms of the Dirichlet projector $Q^\mu_{\,\,\,\nu}$, which projects
vectors onto the space normal to the brane, we have
\beq
R^\mu_{\,\,\,\rho} Q^\rho_{\,\,\,\nu}
  =Q^\mu_{\,\,\,\rho} R^\rho_{\,\,\,\nu}= -Q^\mu_{\,\,\,\nu} \,.
\eeq{RQ}
Correspondingly, the Neumann condition reads
\beq
 N^\rho_{\,\,\,\mu} E_{\sigma\rho} N^\sigma_{\,\,\,\nu}
  -N^\rho_{\,\,\,\mu} E_{\rho\sigma} N^\sigma_{\,\,\,\lambda}
  R^\lambda_{\,\,\,\nu} =0 \,,
\eeq{piEpiR}
where $N^\mu_{\,\,\,\nu} = \delta^\mu_\nu - Q^\mu_{\,\,\,\nu}$
is the Neumann projector, projecting vectors onto the tangent space
of the brane. This condition tells us that for spacefilling D-branes
(i.e., when the Neumann projector is the identity), the
gluing matrix is given by $R=E^{-1} E^T$.
In addition one finds that the metric diagonalises with respect to
the D-brane,
\beq
N^\mu_{\,\,\,\rho} g_{\mu\nu} Q^\nu_{\,\,\,\sigma} =0\,,
\eeq{pigQ}
and that, at least in the presence of an Abelian isometry, the Neumann projector is integrable along the isometry direction,
\beq
N^\mu_{\,\,\,\gamma} N^\rho_{\,\,\,\nu}  N^\delta_{\,\,\,[\mu , \rho]}  =0\,.
\eeq{piinteg1}
The latter condition is essentially the statement that the D-brane is a
well-defined submanifold of the target space \cite{Yano}, and in this paper we shall assume that it holds on $G$.

\subsection{Poisson-Lie T-duality}

We now put the sigma model in the framework necessary for finding
its Poisson-Lie T-dual, explaining how the duality acts on the background field $E_{\mu\nu}$.
The action (\ref{defSX}) may be rewritten in terms of group elements $g\in G$:
$$
S =  \int d^2\xi \,\, L^a_{\+}(g) L^b_{=}(g) E_{ab}(g) \,,
$$
where $a,b$ are Lie algebra indices,
$L^a_{\pp} \equiv L(g)^a_\mu \d_{\pp}X^\mu = (g^{-1}\d_{\pp}g)^a$ is the left-invariant vector field, and 
$$
E_{ab}(g)\equiv (L(g)^{-1})^\mu_aE_{\mu\nu}(g) (L(g)^{-1})^\nu_b \,.
$$
We choose a basis $\{T_a, \wt T^a\}$ of $\D$ such that $\G=\span\{T_a\}$ and
$\wt\G=\span\{\wt T^a\}$, satisfying the following orthogonality conditions,\footnote{The vanishing brackets in (\ref{Torthog}) imply that $\G$ and $\wt \G$ are maximally isotropic in $\D$.}
\beq
\langle T_a \,, \,  \wt T^b \rangle = \delta^b_a\, ,
\quad\quad
\langle T_a \,, \, T_b \rangle =
\langle \wt T^a \,, \,  \wt T^b \rangle = 0 \, .
\eeq{Torthog}
Here $\langle \,\cdot \,, \, \cdot\, \rangle$ is the
non-degenerate bilinear form on $\D$ invariant under the adjoint action
of $D$,
$$
\langle Ad_l \,\,v, Ad_l  \,\,w\rangle = \langle v , w \rangle \,,\,\,\,\,v,w \in\D\,,
\,\,l \in D\,,
$$
where $Ad_l  \,\,v \equiv lv l^{-1}$. The adjoint map
corresponds to the left-action of $D$ on itself in the adjoint representation, since
$$
Ad_k  \,\,Ad_l  \,\,v =  k lv l^{-1} k^{-1} = Ad_{kl} \,\, v\,,\,\,\,\, k,l\in D\,.
$$
We may write the adjoint representation in terms of the matrices $a,b,d$ defined as the coefficients in the expansion
\beq
g^{-1}T_a g \equiv a(g)_a^{\,\,\,b} T_b \,, \qquad
g^{-1} \wt T^a g \equiv b(g)^{ab} T_b + d(g)^a_{\,\,\,b} \wt T^b \,.
\eeq{adrep}

\subsubsection{Canonical transformations}

Klim\v{c}\'{\i}k and \v{S}evera showed \cite{KS1,Klimcik9} how the general background
field $E(g)$
can be found by translating a general
$g$-independent reference field $E(e)$ from the identity $e\in G$ to the point
$g\in G$, by left- or right-action of $G$ on itself. If we use left-translation the result is \cite{vonUnge}
$$
[r(g)^{-1}]^\mu_a E_{\mu\nu}(g) [r(g)^{-1}]^\nu_b =
\left( [a(g)^{-1} + E(e) b(g)^{T}]^{-1}\right)_a^{\,\,\, c} E_{cd}(e) [d(g)^{-1}]^d_{\,\,\, b}\,,
$$
where $[r(g)^{-1}]^\mu_a$ is the inverse of the right-invariant vector field $r(g)^a_\mu$, and the superscript $T$ denotes transpose.
Alternatively, this may be expressed in terms of the natural Poisson
bracket $\Pi(g)$ on $G$ \cite{KS3},
$$
\Pi^{ab} (g) \equiv \langle g^{-1} \wt T^c g , \wt T^a \rangle
\langle g^{-1} T_c g , \wt T^b  \rangle = b(g)^{ca}a(g)_c^{\,\,\,b}\,,
$$
as follows (note that $(d^{-1})^a_{\,\,\,b} = a_b^{\,\,\,a} =(r^{-1})^\mu_b L^a_\mu$, where we simplify notation by dropping the explicit $g$-dependence),
\ber
\nonumber E_{ab}(g) =
(L^{-1})^\mu_a E_{\mu\nu}(g)  (L^{-1})^\nu_b &=&
(a^{-1})_a^{\,\,\, c} (r^{-1})^\mu_c E_{\mu\nu}(g) (r^{-1})^\nu_d
(a^{-1})_b^{\,\,\,d} \\
&=& \left([E(e)^{-1}  +  \Pi(g)]^{-1} \right)_{ab} \,.
\eer{Ee2Eg2}
Similarly, the dual background $\wt E$ can be transported from $\wt e \in \wt G$ to any point $\wt g\in \wt G$ by left-action of $\wt G$ on itself.
Defining the matrices $\wt a,\wt b,\wt d$ as
$$
\wt g^{-1} \wt T^a \wt g \equiv \wt a(\wt g)^a_{\,\,\,b} \wt T^b \,, \qquad
\wt g^{-1} T_a \wt g \equiv \wt b(\wt g)_{ab} \wt T^b +
\wt d(\wt g)_a^{\,\,\,b} T_b \,,
$$
the dual background is found to be
\beq
\wt E^{ab}(\wt g)
=\left([\wt E(\wt e)^{-1}+ \wt \Pi(\wt g)]^{-1} \right)^{ab} \,,
\eeq{wtEe2Eg}
where $\wt E^{ab}(\wt g)\equiv (\wt L^{-1})^a_\mu \wt E^{\mu\nu}(\wt g)  (\wt L^{-1})^b_\nu$, $\wt E(\wt e)=E(e)^{-1}$ (this follows from orthogonality, with respect to
the bilinear form $\langle \,\cdot \,, \, \cdot\, \rangle$, of the graphs
produced by the left-action of $G$ and $\wt G$ respectively \cite{KS1,Klimcik9}),
and
$$
\wt \Pi_{ab} (\wt g) \equiv \langle \wt g^{-1} T_c \wt g , T_a \rangle
\langle \wt g^{-1} \wt T^c  \wt g,  T_b   \rangle
= \wt b(\wt g)_{ca}\wt a(\wt g)^c_{\,\,\,b}
$$
is the natural Poisson bracket on $\wt G$. 
The backgrounds $E(g)$ and $\wt E(\wt g)$ are thus related via $E(e)$,
and this relation is determined by the transformation (\ref{wtEe2Eg}).

Also the worldsheet fields $L^a_{\pp}$ can be dualised directly, by applying a canonical transformation. This transformation was found by Sfetsos \cite{Sfetsos1} and is given by 
\ber
L^a_{\sigma} &=& \left(\delta^a_b - \Pi^{ac} \wt\Pi_{cb} \right) \wt P^b
- \Pi^{ab} (\wt L_{\sigma})_b \,,\label{cantransL} \\
P_a &=& \wt\Pi_{ab} \wt P^b + (\wt L_{\sigma})_a \,,\label{cantransP} 
\eer{cantrans}
where the canonical variables are defined as
\ber
L^a_\sigma &\equiv & \half \left(L^a_{\+}
- L^a_{=}\right) \,, \label{canvarL} \\
P_a  &\equiv & L^\mu_a P_\mu  = L^\mu_a
\frac{\delta \cL}{\delta (\d_{\tau} X^\mu)}
= \half \left(E_{ba} L^b_{\+} + E_{ab} L^b_{=} \right) \,. \label{canvarP} 
\eer{canvar}
Sfetsos showed in \cite{Sfetsos2} that the transformations (\ref{cantransL})
and (\ref{cantransP}) are locally well-defined and that they preserve the form of the Hamiltonian as well as the canonical
Poisson brackets for the conjugate pair of variables $(L^a_\sigma,P_a)$.

\subsubsection{Poisson-Lie condition}

The backgrounds (\ref{Ee2Eg2}) and (\ref{wtEe2Eg}) are solutions of the \emph{Poisson-Lie condition}, a necessary condition for two models to be Poisson-Lie T-dual. It is equivalent to the condition that the worldsheet is horizontally liftable to the double.
For a connected Poisson-Lie group $G$ and its dual Poisson-Lie group $\wt G$, there are homomorphisms of Lie groups $G\hookrightarrow D$ and $\wt G \ \hookrightarrow D$, and one can define a product map $G \times \wt G \rightarrow D$ which is a diffeomorphism onto a neighbourhood of the identity in $D$ \cite{Chari}. If $G$ is compact and the image of $\wt G$ in $D$ is closed, then this map is a diffeomorphism of $G \times \wt G$ onto $D$. An element
$g\in G$ can thus be lifted to the double by multiplying it with an element $\wt h\in\wt G$. This lift can be factorised in two ways:
\beq
l= g \wt h= \wt g h\,,
\eeq{glift}
for some elements $\wt g\in \wt G$ and $h\in G$. Eq.~(\ref{glift}) defines the \emph{dressing action} of $G$ on $\wt G$, whose orbits in $\wt G$ coincide with the symplectic leaves in the stratification of $\wt G$ \cite{Malkin, Chari}. 

Two extremal worldsheets $\Sigma \hookrightarrow G$ and $\wt\Sigma \hookrightarrow \wt G$ are Poisson-Lie T-dual if and only if they
can be horizontally lifted to a surface $\Sigma_D\hookrightarrow D$.
The condition for horizontal
liftability is that the currents $J(g)$ and $\wt J (\wt g)$
associated with the left-translation of, respectively, $\Sigma$ in
$G$ and $\wt \Sigma$ in $\wt G$, are flat connection
one-forms. That is, they must satisfy the zero-curvature conditions
(here $f_{\,\,\,\,bc}^{a}$ and $\wt f^{\,\,\,\,bc}_{a}$ are the structure
constants of $\G$ and $\wt \G$, respectively)
\beq
\begin{array}{r}
dJ_a +\half \wt f^{\,\,\,\,bc}_{a}J_b\wedge J_c=0 \,, \\
d\wt J^a +\half f_{\,\,\,\,bc}^{a}\wt J^b\wedge \wt J^c=0 \,,
\end{array}
\eeq{FlatJ}
the solutions of which may be written as
\beq
J(g)=-d\wt h \,\,\wt h^{-1} \,, \qquad
\wt J (\wt g)=-dh\,\, h^{-1} \,,
\eeq{Jsolns}
where $\wt h$ and $h$ are the auxiliary elements used in the lift (\ref{glift}).
By lifting the worldsheet boundaries of the free open string into the double in this way
Klim\v{c}\'{\i}k and \v{S}evera \cite{KS2} showed that the dual D-branes in $\wt G$ coincide
with the symplectic leaves defined by the associated dressing action.
One can show that eqs.~(\ref{FlatJ}) are equivalent to the conditions (see appendix~\ref{a:Econd})
$$
\begin{array}{rcl}
\cL_{r^a} E_{\mu\nu} (g)&=& -E_{\mu\rho} (g) (r^{-1})^\rho_b \wt f^{\,\,\,\,bc}_{a}
(r^{-1})^\sigma_cE_{\sigma\nu} (g)\,, \\
\cL_{\wt r_a} \wt E^{\mu\nu} (\wt g)&=& -\wt E^{\mu\rho} (\wt g)
(\wt r^{-1})_\rho^b  f_{\,\,\,\,bc}^{a}
(\wt r^{-1})_\sigma^c \wt E^{\sigma\nu} (\wt g)\,.
\end{array}
$$
These are the Poisson-Lie conditions; they are manifestly symmetric under interchange of $G$ and $\wt G$.
They can alternatively be obtained as the equations of
motion of a first order action defined on the double \cite{vonUnge,KS4}, from which
the sigma models on $G$ and $\wt G$ may be derived by inserting the two factorisations (\ref{glift}) and integrating out $\wt h$ or $h$, respectively.

\Section{Boundary conditions}
\label{bconditions}

We now apply the canonical transformations to the worldsheet boundary conditions, to find their Poisson-Lie T-dual counterparts.
In the Lie algebra frame the boundary conditions (\ref{origbc0})--(\ref{piinteg1}) read
\beq
L^a_{=} = R^a_{\,\,\,b} L^b_{\+} \,,
\eeq{origbc}
\beq
R^c_{\,\,\,a} g_{cd} R^d_{\,\,\,b}=g_{ab} \,,
\eeq{RgR}
\beq
R^a_{\,\,\,c} Q^c_{\,\,\,b}  =Q^a_{\,\,\,c} R^c_{\,\,\,b} = -Q^a_{\,\,\,b} \,,
\eeq{LieRQ}
\beq
 N^c_{\,\,\,a}E_{dc} N^d_{\,\,\,b}
  -N^c_{\,\,\,a} E_{cd} N^d_{\,\,\,e}  R^e_{\,\,\,b} =0 \,,
\eeq{LiepiEpiR}
\beq
N^c_{\,\,\,a} g_{cd} Q^d_{\,\,\,b}  =0\,,
\eeq{LiepigQ}
\beq
N^c_{\,\,\,a} N^e_{\,\,\,b} N^d_{\,\,\,[c , e]}  =0\,,
\eeq{Liepiinteg1}
where $R^a_{\,\,\,b} \equiv L^a_\mu  R^\mu_{\,\,\,\nu} (L^{-1})^\nu_b$ and
similarly for $N^a_{\,\,\,b}$ and $Q^a_{\,\,\,b}$.
To find the dual conditions, we use
eqs.~(\ref{Ee2Eg2}), (\ref{wtEe2Eg}), (\ref{canvarL}) and (\ref{canvarP}) to rewrite the canonical transformations
(\ref{cantransL}), (\ref{cantransP}) as acting on $L^a_{\pp}$. If we suppress
indices, the resulting map is
\ber
\wt L_{\+}  &=& (\wt E^T)^{-1} (E_0^T)^{-1} E^T L_{\+}  \,, \label{cantransLp} \\
\wt L_{=}   &=&-\wt E^{-1} E_0^{-1}  E L_{=} \,. \label{cantransLm} 
\eer{cantransLpm}
Note that the transformation from $L_{\pp}$ to $\wt L_{\pp}$ is the
Lie algebra map $\G \rightarrow \wt\G$ corresponding to the map
$G \rightarrow \wt G$ induced by the dressing action (\ref{glift})
of $G$ on $\wt G$.
Under (\ref{cantransLp}), (\ref{cantransLm}) the boundary condition
(\ref{origbc}) transforms to
$$
(\wt L_{=})_a = \wt R_a^{\,\,\,b} (\wt L_{\+})_b\,,
$$
where $\wt R_a^{\,\,\,b} \equiv \wt L_a^\mu \wt  R_\mu^{\,\,\,\nu} (\wt L^{-1})_\nu^b$ and
\beq
\wt R \equiv - \wt E^{-1} E_0^{-1} E R (E^T)^{-1} E_0^T \wt E^T \,.
\eeq{Rtilde}
This is the transformation of the gluing matrix which defines how Poisson-Lie T-duality acts on the sigma model boundary conditions.

The dual of the conformality condition (\ref{RgR}) is found to be
$$
\wt R_c^{\,\,\,a} \wt g^{cd} \wt R_d^{\,\,\,b}=\wt g^{ab}\,,
$$
where the transformation of the metric,
\beq
\wt g= \wt E E_0 E^{-1} g (E^T)^{-1} E_0^T  \wt E^T \,,
\eeq{dualmetric}
follows from (\ref{Ee2Eg2}) and (\ref{wtEe2Eg}).
Hence conformal symmetry is automatically satisfied on the
dual side. Furthermore, the transformation law (\ref{Rtilde}) for the gluing matrix determines also the form of the dual Neumann and Dirichlet projectors $\wt N$ and $\wt Q$, via the defining relation $\wt R \wt Q = \wt Q \wt R = -\wt Q$. One may thus consistently impose the dual versions of conditions (\ref{LiepiEpiR})-(\ref{Liepiinteg1}) to obtain a well-defined open string theory on the dual target manifold.

The duality transformations (\ref{cantransLp}), (\ref{cantransLm}), (\ref{Rtilde}) and (\ref{dualmetric}) may be interpreted as a generalisation of the Abelian T-duality discussion in \cite{ALZ4,Hassan} if we define
$$
\Qp  \equiv (\wt E^T)^{-1}  (E^T_0)^{-1}  E^T \,,\qquad
\Qm \equiv -\wt E^{-1} E_0^{-1} E \,.
$$
Then we have
$$
\wt L_{\+} = \Qp L_{\+} \,, \qquad
\wt L_{=} = \Qm L_{=} \,,
$$
$$
\wt R = \Qm R \Qp^{-1}\,, \qquad
\wt g= (\Qp^T)^{-1} \, g \, \Qp^{-1}\,,
$$
i.e., precisely the relations listed in \cite{ALZ4} for the Abelian case, but written in the Lie algebra
frame.

Before proceeding to discuss specific examples, we highlight some general features of the gluing matrix $R$ and its dual $\wt R$.
First note that $\det(\wt{R})=\det(-R)$,
a result that will be useful in the case-by-case analysis
in the next section. Next we see that,
in coordinates adapted to the D-brane, $R$ takes the form
$$
R = \left(\begin{array}{cc}
E_N^{-1} E_N^T &0 \\
0&-\bid
\end{array}\right) \,,
$$
where $E_N$ is the nonzero Neumann-Neumann block in the matrix $N^T EN$.
If the B-field restricts to zero on the brane (i.e., $B_{ab}$ has no Neumann-Neumann part), so that $E_N$ is symmetric, then the gluing matrix is simply
$$
R = \left(\begin{array}{cc}
\bid &0 \\
0&-\bid
\end{array}\right)\,.
$$
In this case it can be used to define the Neumann and Dirichlet projectors
via $N=(\bid + R)/2$ and $Q=(\bid - R)/2$, since $R^2= \bid$ \cite{ALZ1}.
In general, however, this is not the case, and we have $R^2\neq \bid$.

\Section{Examples}
\label{examples}

There are three types of Drinfel'd double, depending on whether or not the
two constituent Poisson-Lie groups $G$ and $\wt G$ are Abelian. Here we consider each type in turn, analysing the consequences of the duality transformation (\ref{Rtilde}) for specific examples.

\subsection{Abelian double}

Take the Drinfel'd double of $D=U(1)^n\times U(1)^n$, where Poisson-Lie T-duality is expected to reduce to traditional Abelian T-duality. The Poisson bracket vanishes, and the Poisson-Lie condition is just the isometry condition $\cL_{r^a}E_{\mu\nu}=0$. We have $E=E_0=\wt E_0^{-1} = \wt E^{-1}$, so $E$ and $\wt E$ are both independent of the group elements. From eqs.~(\ref{LieRQ}) and (\ref{LiepiEpiR}) we see that $R$ is constant and in adapted
coordinates takes the form
$$
R = \left(\begin{array}{cc}
E_N^{-1} E_N^T &0 \\
0&-\bid
\end{array}\right) \,,
$$
where $E_N$ is the nonzero Neumann-Neumann block in the matrix $N^T E_0N$.
The transformation (\ref{Rtilde}) yields
\beq
\wt R = - E_0 R (E_0^T)^{-1} \,.
\eeq{Rabel}
When $R$ is completely Neumann, i.e., when we have a spacefilling D-brane,
eq.~(\ref{LiepiEpiR}) yields $R=E_0^{-1} E_0^T$ so that (\ref{Rabel}) reduces to
$\wt R = - \bid$. That is, a spacefilling D-brane is T-dual to a pointlike D-brane.
For the model with $G=U(1)$ isometry, this is just the statement that dualisation along the isometry direction transforms it from a Neumann to a Dirichlet direction for the brane. In this case
the transformations (\ref{cantransLp}), (\ref{cantransLm}) reduce to the
familiar maps for the isometry direction $X^0$ (in tangent space),
$$
 \d_\+\wt X^0 =E_0^T \d_\+X^0 \,, \qquad
 \d_= \wt X^0 =-E_0 \d_= X^0\,.
$$
Similarly if the model has $G=U(1)^n$ isometry, the duality map corresponds to simultaneous dualisation of a spacefilling brane along all $n$ directions.

\subsection{Semi-Abelian double}

The semi-Abelian double corresponds to the standard non-Abelian T-duality between a $G$-isometric sigma model with target $G$ and a non-isometric sigma model with the target $\wt\G$ viewed as the Abelian group.
Consider the double of $D=G\ltimes U(1)^n$ (semi-direct product), where $G$ is a non-Abelian group.
An example is the six-dimensional group $D=ISO(3)=SO(3)\ltimes U(1)^3$. 
When $R$ is completely Neumann, then eq.~(\ref{LiepiEpiR}) yields $R=E^{-1} E^T$ and
the duality map (\ref{Rtilde}) reduces to
$$
\wt R = - \wt E^{-1} E_0^{-1} E_0^T \wt E^T =- \wt E^{-1} R_0\wt E^T\,,
$$
where we have defined $R_0\equiv E_0^{-1} E_0^T$.
We moreover have $\wt E= E_0^{-1}$ \cite{Hlavaty1}, hence
$\wt R = - \bid$, so again a spacefilling brane on $G$ is dual to a pointlike brane on the dual manifold.

\subsection{Non-Abelian double}

\subsubsection{Two-dimensional example}

We now turn to a detailed study of the simplest non-Abelian double, namely the Borelian double $gl(2,\bR)$. This is much too simple a model to be physically interesting, but it serves as a tractable toy model to illustrate the main features of the duality transformation. The dual sigma models on this double have been explicitly worked out in \cite{Hlavaty1,Klimcik9}; here we analyse their worldsheet boundary conditions.

We choose the decomposition of the bialgebra as $gl(2,\bR)= (\G,\wt\G)$ with basis $\{T_a\}$ for $\G$ and basis $\{\wt T^a\}$ for $\wt\G$, defined as
$$
T_1 = \left(\begin{array}{cc}
1&0 \\
0&0
\end{array}\right) \,, \qquad
T_2 = \left(\begin{array}{cc}
0&1 \\
0&0
\end{array}\right)\,,
$$
$$
\wt T^1 = \left(\begin{array}{cc}
0&0 \\
0&1
\end{array}\right) \,, \qquad
\wt T^2 = \left(\begin{array}{cc}
0&0 \\
-1&0
\end{array}\right)\,.
$$
The generators $T_a$ span the Borelian subalgebra of
upper triangular matrices and $\wt T^a$ span the Borelian subalgebra of
lower triangular matrices.
Using the parameterisation
$$
g = \left(\begin{array}{cc}
e^\chi &\theta \\
0&1
\end{array}\right)
$$
of group elements $g\in G$, the adjoint representation matrices defined in (\ref{adrep}) read
$$
a(g) = \left(\begin{array}{cc}
1 & \theta e^{-\chi} \\
0 & e^{-\chi} 
\end{array}\right) \,, \qquad
b(g) = \left(\begin{array}{cc}
0 & -\theta e^{-\chi} \\
\theta & \theta^2 e^{-\chi} 
\end{array}\right) \,, \qquad
d(g) = \left(\begin{array}{cc}
1 & 0 \\
-\theta  & e^{\chi} 
\end{array}\right) \,.
$$
If we define the value of the background field $E_{ab}$ at the identity to be the constant matrix
$$
E_0^{-1} = \left(\begin{array}{cc}
x &y \\
u&v
\end{array}\right)\,,
$$
then eq.~(\ref{Ee2Eg2}) yields\footnote{Note that the explicit expressions for
$E(g)$ and $\wt E(\wt g)$ differ from those of \cite{Hlavaty1,Klimcik9}, because they use
right-translation on $G,\wt G$ while we use left-translation, and also we have used slightly different definitions of the fields.}
$$
E(g) =
[(xv-uy)+ \theta e^{-\chi} (\theta e^{-\chi} +y-u)]^{-1}
\left(\begin{array}{cc}
v & -y-\theta e^{-\chi}\\
-u + \theta e^{-\chi}  & x 
\end{array}\right)\,,
$$
where the notation is $E_{11}\equiv E_{\chi\chi}$ etc.
Parameterising the dual group element $\wt g\in \wt G$ as
$$
\wt g = \left(\begin{array}{cc}
1 &0 \\
-\rho & e^\sigma
\end{array}\right)\,,
$$
the corresponding adjoint representations read
$$
\wt a(\wt g) = \left(\begin{array}{cc}
1 & \rho e^{-\sigma} \\
0 & e^{-\sigma} 
\end{array}\right) \,, \qquad
\wt b(\wt g) = \left(\begin{array}{cc}
0 & -\rho e^{-\sigma} \\
\rho & \rho^2 e^{-\sigma} 
\end{array}\right) \,, \qquad
\wt d(\wt g) = \left(\begin{array}{cc}
1 & 0 \\
-\rho  & e^{\sigma} 
\end{array}\right) \,,
$$
and the dual background follows from eq.~(\ref{wtEe2Eg}) (where
$\wt E^{11}\equiv \wt E^{\sigma\sigma}$ etc.),
\begin{eqnarray*}
\wt E(\wt g) &=&
[1+ \rho^2 e^{-2 \sigma}(xv-yu) + \rho e^{-\sigma} (u-y)]^{-1} \times \\
&&\times
\left(\begin{array}{cc}
x & y-\rho e^{-\sigma} (xv-yu)  \\
u+\rho e^{-\sigma} (xv-yu) & v
\end{array}\right)\,.
\end{eqnarray*}
Inserting $E_0$, $E(g)$ and $\wt E(\wt g)$ into eq.~(\ref{Rtilde}),
one finds the dual gluing matrix $\wt R$ for any given original gluing matrix $R$.
There are three different possibilities:

\noindent\emph{Case 1:}
$$
R = \left(\begin{array}{rr}
-1 &  0 \\
 0 & -1
\end{array}\right)
$$
This is a D(-1)-brane, with Dirichlet directions in both
coordinate directions on $G$. Then eq.~(\ref{Rtilde}) yields
$$
\wt R \equiv \wt E^{-1} E_0^{-1} E (E^T)^{-1} E_0^T \wt E^T
\equiv -\frac{1}{AB}\left(\begin{array}{cc}
\wt R_{11} &\wt R_{12}\\
\wt R_{21} &\wt R_{22}
\end{array}\right)
$$
where
$$
A\equiv xv-uy+\theta e^{-\chi}(\theta e^{-\chi}+y-u) , \qquad
B\equiv 1+\rho e^{-\sigma} (u-y)+\rho^2 e^{-2\sigma} (x v-uy)\,,
$$
and
$$
\begin{array}{lcl}
\wt R_{11} &=& [(y+\theta e^{-\chi}) (1+u\rho e^{-\sigma})-xv\rho e^{-\sigma}]^2-xv(1+\rho e^{-\sigma}\theta e^{-\chi})^2 \\
\wt R_{12} &=& v (1+\rho e^{-\sigma}\theta e^{-\chi}) [(y+\theta e^{-\chi}) (1+u\rho e^{-\sigma})+(-u+\theta e^{-\chi}) (1-y\rho e^{-\sigma})\\
&&-2xv\rho e^{-\sigma} ] \\
\wt R_{21} &=&  -x (1+\rho e^{-\sigma}\theta e^{-\chi}) [(y+\theta e^{-\chi}) (1+u\rho e^{-\sigma})+(-u+\theta e^{-\chi}) (1-y\rho e^{-\sigma}) \\
&& -2xv\rho e^{-\sigma}] \\
\wt R_{22}&=& [(-u+\theta e^{-\chi}) (1-y\rho e^{-\sigma})-xv\rho e^{-\sigma}]^2-xv(1+\rho e^{-\sigma}\theta e^{-\chi})^2
\end{array}
$$
$\wt R$ is a nontrivial matrix in general. Its determinant is
$\det \wt R = \det (-R) = 1$, so the dual brane has either zero or two Dirichlet directions. If the latter, then the only solution is $\wt R=-\bid$, which happens only for backgrounds $E(g)$ and $\wt E(\wt g)$ such that
$ \wt E (\wt E^T)^{-1} = - E_0^{-1} E (E^T)^{-1} E_0^T$. If the dual brane has zero Dirichlet directions, then it is a D1-brane, whose embedding in $\wt G$ is given by $\wt R$. This situation occurs only if the Poisson bracket $\Pi$ on $G$ vanishes\footnote{We are grateful to Libor \v{S}nobl for this observation. In
dimensions higher than two, the condition is $\det \Pi =0$.},
since in this case the relation (\ref{Rtilde}) reduces to $E_0^{-1} E (E^T)^{-1} E_0^T=\bid$, implying $\Pi(E_0+E_0^T)=0$, and hence (since $E_0+E_0^T=0$ would imply a vanishing metric) we find $\Pi=0$. We conclude that the D(-1)-brane is dual either to a D1-brane (provided $\Pi=0$), or possibly, for some special backgrounds, a D(-1)-brane.

\noindent\emph{Case 2:}
\beq
R = \left(\begin{array}{rr}
1 &  0 \\
 0 & -1
\end{array}\right)
\eeq{R1_1}
This is a D0-brane, with one Dirichlet direction and one
Neumann direction. The dual gluing matrix again follows from eq.~(\ref{Rtilde}):
$$
\wt R
\equiv -\frac{1}{AB}\left(\begin{array}{cc}
\wt R_{11} &\wt R_{12}\\
\wt R_{21} &\wt R_{22}
\end{array}\right)
$$
where
$$
\begin{array}{lcl}
\wt R_{11} &=& [(y+\theta e^{-\chi}) (1+u\rho e^{-\sigma})-xv\rho e^{-\sigma}]^2+xv(1+\rho e^{-\sigma}\theta e^{-\chi})^2 \\
\wt R_{12} &=& v(u+y)(1+\rho e^{-\sigma}\theta e^{-\chi})^2\\
\wt R_{21} &=& -x(u+y)(1+\rho e^{-\sigma}\theta e^{-\chi})^2\\
\wt R_{22}&=& -\{[(-u+\theta e^{-\chi}) (1-y\rho e^{-\sigma})-xv\rho e^{-\sigma}]^2+xv(1+\rho e^{-\sigma}\theta e^{-\chi})^2\}
\end{array}
$$
The determinant
is $\det \wt R =  -1$, so $\wt R$ has one $+1$ eigenvalue
and one $-1$ eigenvalue. This means $\wt R$ can be diagonalised
to take the form (\ref{R1_1}). Hence the dual D-brane also
has one Dirichlet direction and one Neumann direction,
so the dual of the D0-brane is a D0-brane.
The original D0-brane lies along one of the coordinate directions in
the original manifold whereas the dual D0-brane is nontrivially embedded in the dual manifold, and the embedding can be found explicitly
by diagonalising $\wt R$. As a special case, note that if $E_0=\bid$,
then $\wt R = \diag (-1,1)$, i.e., Neumann and Dirichlet directions are
just swapped relative to the original brane.

\noindent\emph{Case 3:}
$$
R = \left(\begin{array}{rr}
\alpha    &  \beta \\
\gamma &  \delta
\end{array}\right)
$$
For generic eigenvalues, this is a D1-brane (i.e., spacefilling), which according to eq.~(\ref{LiepiEpiR}) is given by
$$
R=E^{-1}E^T =
\frac{1}{A}\left(\begin{array}{cc}
xv - (y+\theta e^{-\chi})^2   &  -x(u-y - 2\theta e^{-\chi})   \\
v(u-y - 2\theta e^{-\chi})& xv - (u-\theta e^{-\chi})^2
\end{array}\right) 
$$
The dual matrix becomes
$$
\wt R =
 -\frac{1}{B}\left(\begin{array}{cc}
(1+u\rho e^{-\sigma})^2-xv\rho^2 e^{-2\sigma}
&v\rho e^{-\sigma}(2+(u-y)\rho e^{-\sigma})\\
-x\rho e^{-\sigma}(2+(u-y)\rho e^{-\sigma})
&(1-y\rho e^{-\sigma})^2-xv\rho^2 e^{-2\sigma}
\end{array}\right)
$$
The determinant is
$\det \wt R =1$, so the dual brane has either zero or two Dirichlet directions. If it has two Dirichlet directions, then we obtain exactly the reverse situation of Case 1: the D1-brane is dual to a D(-1)-brane provided the Poisson bracket $\wt \Pi$ on $\wt G$ vanishes, and we have $\wt R = -\bid$.
If on the other hand the dual brane has zero Dirichlet directions, then it is a
D1-brane, and since it is spacefilling it should satisfy the dual version
of eq.~(\ref{LiepiEpiR}), $\wt R =\wt E^{-1} \wt E^T$. It turns out, however, that this situation is disallowed by eq.~(\ref{Rtilde}), because it would require $E_0+E_0^T=0$ and hence a vanishing metric.
We conclude that D1-branes are dual to D(-1)-branes provided $\wt \Pi =0$,
but that D1-branes are never dual to D1-branes.

The Borelian example nicely illustrates the symmetric nature of 
Poisson-Lie T-duality. The transformation law (\ref{Rtilde}) for the gluing matrix is completely reversible, on the one hand interchanging D(-1)- and D1-branes independently of which of the two types of brane one starts with, and on the other hand taking D(-1)-branes to D(-1)-branes and D0-branes to D0-branes. It is moreover manifestly symmetric under interchange of the two groups $G$ and $\wt G$ corresponding to the Drinfel'd double.

\subsubsection{Three-dimensional example}

We also work out an example where the target spaces are three-dimensional,
namely the double studied by Sfetsos in \cite{Sfetsos3}. The algebras in this
double are $\G=su(2)$ and $\wt\G=e_3$, whose generators $T_a$ and $\wt T^a$, respectively, satisfy the commutation relations ($a,b=(i,3), i=1,2$)
$$
[T_a,T_b] = i\epsilon_{abc}T_c\,, \qquad
[\wt T^3,\wt T^i] = \wt T^i\,, \qquad
[\wt T^i,\wt T^j] =0\,,
$$
$$
[T_i,\wt T^j] =i\epsilon_{ij} \wt T^3 - \delta_{ij} T_3\,, \qquad
[T_3,\wt T^i] = i\epsilon_{ij} \wt T^j\,, \qquad
[\wt T^3, T_i] =  i\epsilon_{ij} \wt T^j - T_i\,.
$$
Adopting the notation and assumptions of Sfetsos, we define the constant background
at the identity as $E_0^{-1} = \diag(\lambda_1,\lambda_2,\lambda_3)$
with $\lambda_2 \equiv \lambda_1$, and the Poisson brackets obtained
from $\Pi = b^T a, \wt \Pi = \wt b^T \wt a$ (for explicit expressions for matrices $a, b$ etc, see ref.~\cite{Sfetsos3}) may be written in terms of the components $A_a, \wt A_a$ of three-vectors $\vec{A}, \vec{\wt A}$ as
$$
\Pi^{ab} = -\epsilon_{abc} A_c \,, \qquad
\wt \Pi_{ab} = -\epsilon_{abc} \wt A_c \,,
$$
where, in terms of local coordinates $(\psi,\theta,\phi)$ on $\G$
and  $(y_1,y_2,\chi)$ on $\wt\G$,
$$
\vec A \equiv (\cos \psi \sin \theta,\sin\psi \sin\theta,\cos\theta -1)\,, 
$$
$$
\vec{\wt A} \equiv (y_1 e^{-\chi},y_2 e^{-\chi},\sinh \chi\,\, e^{-\chi}-\half (y_1^2 + y_2^2) e^{-2\chi}) \,.
$$
Then the background fields $E,\wt E$ read
$$
E_{ab} = \frac{1}{V} \left(\begin{array}{ccc}
\lambda_1 \lambda_3 +A_1^2 & \lambda_3 A_3 + A_1 A_2 
& -\lambda_1 A_2 + A_1 A_3 \\
-\lambda_3 A_3 + A_1 A_2 & \lambda_1 \lambda_3 +A_2^2 & \lambda_1 A_1 + A_2 A_3 \\
\lambda_1 A_2 + A_1 A_3  & -\lambda_1 A_1 + A_3 A_2 &\lambda_1^2 +A_3^2
\end{array}\right)\,,
$$
$$
\wt E^{ab} = \frac{1}{\wt V} \left(\begin{array}{ccc}
\lambda_1 (1+\lambda_1\lambda_3 \wt A_1^2) &
\lambda_1^2 (\wt A_3+\lambda_3 \wt A_1\wt A_2) &
\lambda_1\lambda_3 (-\wt A_2+\lambda_1 \wt A_1\wt A_3)  \\
\lambda_1^2 (-\wt A_3+\lambda_3 \wt A_1\wt A_2)  &
\lambda_1 (1+\lambda_1\lambda_3 \wt A_2^2) &
\lambda_1\lambda_3 (\wt A_1+\lambda_1 \wt A_2\wt A_3)   \\
\lambda_1\lambda_3 (\wt A_2+\lambda_1 \wt A_1\wt A_3)    &
\lambda_1\lambda_3 (-\wt A_1+\lambda_1 \wt A_2\wt A_3) &
\lambda_3 (1+\lambda_1^2 \wt A_3^2)
\end{array}\right)\,,
$$
where
$$
V\equiv   \lambda_1^2 \lambda_3 + \lambda_1 A_1^2
+ \lambda_1 A_2^2+ \lambda_3 A_3^2\,,\qquad
\wt V \equiv  1+ \lambda_1\lambda_3\wt A_1^2+ \lambda_1\lambda_3\wt A_2^2+ \lambda_1^2\wt A_3^2 \,.
$$

In a three-dimensional manifold we can have four different types of D-brane:
D(-1), D0, D1, and D2. We compute the dual gluing matrix for each of these cases.

\noindent\emph{Case 1:} $R = -\bid$, a D(-1)-brane.
Then eq.~(\ref{Rtilde}) yields the dual gluing matrix
$$
\begin{array}{l}
(\wt R)_a^{\,\,\,b} = -\frac{1}{\wt V} \sum_{c,d,e,f,g}
\left(\delta_{ad} - \epsilon_{adc} \lambda_d \wt A_c \right)  \times \\
 \times 
\left(\delta_{de} [1-2\lambda_1^2 \lambda_3/V] - 
2\lambda_e[\epsilon_{def} \lambda_f A_f +A_d A_e]/V \right)\left( \delta_{eb} - \epsilon_{ebg} \lambda_b \wt A_g
+   \wt A_e \wt A_ b \lambda_1^2 \lambda_3 /\lambda_e \right) \,.
\end{array}
$$
It has determinant $\det \wt R = \det(-R) = 1$, so it can have
zero or two Dirichlet directions, i.e., it is either a D2-brane (spacefilling) or a
D0-brane. Note that, while in two dimensions the condition for a D(-1)-brane to be dual to a spacefilling brane is that $\Pi=0$, the corresponding condition
in higher dimensions is the less restrictive $\det \Pi =0$. Since in three dimensions this is always true,
there is a priori no obstruction for the D(-1)-brane to be dual to a D2-brane.

\noindent\emph{Case 2:} $R=\diag(1,-1,-1)$, a D0-brane.
The dual gluing matrix reads
$$
\begin{array}{l}
(\wt R)_a^{\,\,\,b} = -\frac{1}{\wt V} \sum_{c,d,e,f,g}
\left(\delta_{ad} - \epsilon_{adc} \lambda_d \wt A_c \right) \left[
\delta_{d1} \delta_{e1} + (\delta_{d2} \delta_{e2} +\delta_{d3} \delta_{e3} )
(-1+2\lambda_1A_1^2/V) \right. \\
 \left.
+(\delta_{d2} \delta_{e2} -\delta_{d3} \delta_{e3} )2A_1A_2A_3/V
- 2(1-\delta_{d1})(1-\delta_{e1})\epsilon_{def}(A_d^2+\lambda_e\lambda_f)A_f/V
\right. \\
 \left.
+2\delta_{d1} (1-\delta_{e1}) A_1(\epsilon_{1ef}A_1A_f-\lambda_e A_e)/V
 \right]  \left( \delta_{eb} - \epsilon_{ebg} \lambda_b \wt A_g
+   \wt A_e \wt A_ b \lambda_1^2 \lambda_3 /\lambda_e \right) \,.
\end{array}
$$
The determinant is $-1$, i.e., it is either a D(-1)-brane or a D1-brane.

\noindent\emph{Case 3:} $R=\diag(R_N,-1)$, a D1-brane
where the submatrix
$$
R_N = \left(\begin{array}{cc}
1-2(\lambda_3A_3^2 +A_1 A_2 A_3)/V &
-2A_3(\lambda_1\lambda_3 + A_2^2)/V  \\
2A_3(\lambda_1\lambda_3 + A_1^2)/V&
1-2(\lambda_3A_3^2 -A_1 A_2 A_3)/V
\end{array}\right)
$$
is determined by\footnote{The inverse here is understood to be taken on the
Neumann subspace.}
$R_N = (N^TEN)^{-1}(N^TE^TN)$, with
the Neumann projector $N=\diag(1,1,0)$.
The dual gluing matrix reads
$$
\begin{array}{l}
(\wt R)_a^{\,\,\,b} = -\frac{1}{\wt V} \sum_{c,d,e,f,g}
\left(\delta_{ad} - \epsilon_{adc} \lambda_d \wt A_c \right)  
\left[(1-\delta_{d3})\left\{(1-\delta_{e3}) \right. \right. \times \\
\times \left. \left. [\delta_{de}
(1+2(1-2\delta_{d1})A_1A_2A_3/V) +2\epsilon_{def}A_d^2 A_f/V]
-2 \delta_{e3} \lambda_3 A_3 A_d/V \right\} -\delta_{d3}\delta_{e3}\right]
\times \\
\times
\left( \delta_{eb} - \epsilon_{ebg} \lambda_b \wt A_g
+   \wt A_e \wt A_ b \lambda_1^2 \lambda_3 /\lambda_e \right) \,.
\end{array}
$$
Its determinant is $1$, so we have a D0-brane or a D2-brane.

\noindent\emph{Case 4:} $R=E^{-1} E^T$, a D2-brane.
The dual gluing matrix reads
$$
\begin{array}{l}
(\wt R)_a^{\,\,\,b} =
\delta_{ab}- 2(
\delta_{ab} - \sum_{c} \epsilon_{abc} \lambda_b \wt A_c
+   \wt A_a \wt A_b \lambda_1^2 \lambda_3 /\lambda_a
)/{\wt V} \,.
\end{array}
$$
The determinant is $-1$, so it is a D(-1)-brane or a D1-brane.
As in Case 1, note that since $\det \wt \Pi =0$, the D2-brane can be
dual to a D(-1)-brane.

To summarise Poisson-Lie T-duality in this three-dimensional example,
the D-branes in the model are exchanged as follows:
$$
\begin{array}{ccc}
D(-1) &\leftrightarrow & D0 \\
D0 &\leftrightarrow & D1 \\
D1 &\leftrightarrow & D2 \\
D2 &\leftrightarrow & D(-1)
\end{array}
$$
We see that all branes are linked together in a duality chain, where each step changes the brane dimension by one, except in the duality D(-1) $\leftrightarrow$ D2.

The above analysis is somewhat superficial, considering only
the value of the determinant of the gluing matrix. To obtain more
detailed information about the dual D-branes, one should
study the eigenvalues of each gluing matrix
as well as its explicit form in terms of local coordinates.
In particular, the condition that $E,\wt E$ satisfy eq.~(\ref{Rtilde}) may
for some of the D-brane exchanges
impose restrictions on the variables used in the parameterisation.

\Section{Conclusions}
\label{conclusions}

By applying the Poisson-Lie T-duality canonical transformations found by Sfetsos \cite{Sfetsos1} to the worldsheet boundary conditions of the bosonic nonlinear sigma model, we have derived the explicit duality map, eq.~(\ref{Rtilde}), for the gluing matrix which locally
defines the properties of the D-brane.
The gluing matrix relates left- and right-moving fields on the worldsheet, and the boundary conditions of the open string sigma model are expressed in terms of it. Its eigenvalues determine the dimensionality of the brane, and its form the embedding of the brane in the target space, at least locally. The sigma model and its dual are defined on Poisson-Lie group manifolds that make up a Drinfel'd double, in line with the formalism of Klim\v{c}\'{\i}k and \v{S}evera \cite{KS1}.
We have demonstrated how the boundary conditions transform under Poisson-Lie T-duality, and in particular that the model dual to a conformal model is itself automatically conformal. In the process we had to rewrite the canonical transformations of Sfetsos as a map acting on the relevant worldsheet fields in the Lie algebra frame. It can be written as a direct generalisation of the traditional Abelian T-duality map. We moreover explicitly worked out the duality transformation for the simplest non-Abelian Drinfel'd double, $gl(2,\bR)$, showing how the gluing matrix, and hence the D-brane, transforms under the duality in this case. We found that D0-branes are dual to D0-branes (with different embeddings
in the two dual target spaces), and that, depending on the background fields $E$ and $\wt E$, D(-1)-branes are dual to D(-1)-branes or to D1-branes. This toy model demonstrates the symmetric (or invertible) nature of Poisson-Lie T-duality. We analysed also the three-dimensional double of Sfetsos \cite{Sfetsos3}, finding a similar symmetric duality action on the branes that links
all branes together in a duality chain, where each step changes the brane dimension by one, except in the duality D(-1) $\leftrightarrow$ D2.

The continuation of this programme includes a quest for better geometric understanding of the duality transformation of the gluing matrix, in terms of D-branes in the Drinfel'd double. In particular, the dual gluing matrix in some cases appears to depend on the coordinates of the original manifold, which might indicate a need to restrict the duality to act only on certain types of D-brane. Also, the interpretation of the transformation in terms of Poisson structures and the geometry of symplectic leaves needs clarification.
Extending the analysis to $\N$=1 worldsheet supersymmetric sigma models is an obvious path of investigation, as is a study of the analogous aspects of Poisson-Lie T-plurality \cite{vonUnge2}. In the latter case there exists more than one
maximally isotropic decomposition (Manin triple) of the double into two subalgebras, and the duality transformation must include the switch between decompositions.

\bigskip

\bigskip

{\bf Acknowledgements}:
We are grateful to Ladislav Hlavat\'{y}, Ctirad Klim\v{c}\'{\i}k, Timothy Logvinenko, Libor \v{S}nobl and Rikard von Unge for useful discussions and comments.
CA acknowledges support in part by Deutsche Forschungsgemeinschaft (DFG)
and by the Japanese Society for the Promotion of Science (JSPS). RAR
wishes to thank the staff at ICTP, Trieste for their hospitality during the final stages of this project.

\appendix

\Section{The Poisson-Lie condition}
\label{a:Econd}

The field equations of the action (\ref{defSX})
associated to left-translation on $G$ read
\beq
\begin{array}{r}
\d_{\+} J_{-a} +\d_{=} J_{+a} 
- \cL_{r^a} E_{\mu\nu} \d_{\+} X^\mu \d_{=} X^\nu=0 \,,
\end{array}
\eeq{Econd1}
where the currents $J_{\pm a}$ are defined as
$$
J_{+a}(g) \equiv \d_{\+} X^\mu E_{\mu\nu} (g) (r^{-1})^\nu_a \,, \qquad
J_{-a} (g)\equiv (r^{-1})^\mu_a E_{\mu\nu} (g) \d_{=} X^\nu \,.
$$
To turn (\ref{Econd1}) into a flatness condition for $J_{\pm a}$, we need to impose the following restriction on the background,
\beq
\begin{array}{rcl}
\cL_{r^a} E_{\mu\nu} (g)&=& -E_{\mu\rho} (g) (r^{-1})^\rho_b \wt f^{\,\,\,\,bc}_{a}
(r^{-1})^\sigma_c E_{\sigma\nu} (g)\,,
\end{array}
\eeq{Econd2}
which transforms (\ref{Econd1}) into
\beq
\begin{array}{r}
\d_{\+} J_{-a} +\d_{=} J_{+a} + J_{+b} \wt f^{\,\,\,\,bc}_{a}J_{-c}  =0\,,
\end{array}
\eeq{Econd3}
i.e., precisely the flatness condition (\ref{FlatJ}).


\begin{thebibliography}{6666}
%
%\cite{Ooguri}
\bibitem{Ooguri}
H.~Ooguri, Y.~Oz and Z.~Yin,
``D-branes on Calabi-Yau spaces and their mirrors,''
Nucl.\ Phys.~B {\bf 477} (1996) 407,
hep-th/9606112
%%CITATION = HEP-TH 9606112;%%
%
\bibitem{BorlafLozano}
J.~Borlaf and Y.~Lozano,
``Aspects of T-duality in open strings,''
Nucl.\ Phys.~B {\bf 480} (1996) 239-264,
hep-th/9607051
%%CITATION = HEP-TH 9607051;%%
%
\bibitem{Hori}
K.~Hori,
``Linear models of supersymmetric D-branes,''
hep-th/0012179
%%CITATION = HEP-TH 0012179;%%
%
\bibitem{HKLR86}
C.~M.~Hull, A.~Karlhede, U.~Lindstr\"om and M.~Ro\v{c}ek,
``Nonlinear $\sigma$-models
 and their gauging in and out of superspace,''
Nucl.\ Phys.~B {\bf 266} (1986) 1-44
%
\bibitem{Rocek}
M.~Ro\v{c}ek and E.~Verlinde,
``Duality, quotients, and currents,''
Nucl.\ Phys.~B {\bf 373} (1992) 630-646,
hep-th/9110053
%%CITATION = HEP-TH 9110053;%%
%
\bibitem{Hassan2}
S.~F.~Hassan,
``O(d,d;R) deformations of complex structures and
extended worldsheet supersymmetry,''
Nucl.\ Phys.~B {\bf 454} (1995) 86-102,
hep-th/9408060
%%CITATION = HEP-TH 9408060;%%
%
\bibitem{Alvarez}
E.~Alvarez, J.~L.~F.~Barb\'on and J.~Borlaf,
``T-duality for open strings,''
Nucl.\ Phys.~B {\bf 479} (1996) 218-242,
hep-th/9603089
%%CITATION = HEP-TH 9603089;%%
%
\bibitem{Dorn}
H.~Dorn and H.-J.~Otto,
``On T-duality for open strings in general abelian and nonabelian gauge field backgrounds,''
Phys.~Lett.~B {\bf 381} (1996) 81-88,
hep-th/9603186
%%CITATION = HEP-TH 9603186;%%
%
\bibitem{Forste}
S.~F\"orste, A.~Kehagias and S.~Schwager,
``Non-Abelian duality for open strings,''
Nucl.\ Phys.~B {\bf 478} (1996) 141-155,
hep-th/9604013
%%CITATION = HEP-TH 9604013;%%
%
%\cite{DeWit}
\bibitem{DeWit}
B.~de~Wit, C.~M.~Hull and M.~Ro\v{c}ek,
``New topological terms in gauge invariant actions,''
Phys.\ Lett.~B {\bf 184} (1987) 233-238
%
\bibitem{Giveon}
A.~Giveon and M.~Ro\v{c}ek,
``On nonabelian duality,''
Nucl.\ Phys.~B {\bf 421} (1994) 173-190,
hep-th/9308154
%%CITATION = HEP-TH 9308154;%%
%
\bibitem{KS1}
C.~Klim\v{c}\'{\i}k and P.~\v{S}evera,
``Dual non-abelian duality and the Drinfeld double,''
Phys.\ Lett.~B {\bf 351} (1995) 455-462,
hep-th/9502122
%%CITATION = HEP-TH 9502122;%%
%
\bibitem{STS}
M.~A.~Semenov-Tian-Shansky,
``Dressing transformations and Poisson group actions,"
Publ.\ Res.\ Inst.\ Math.\ Sci.\ Kyoto {\bf 21} (1985) 1237-1260
%
\bibitem{Malkin}
A.~Yu.~Alekseev and A.~Z.~Malkin,
``Symplectic structures associated to Lie-Poisson groups,''
Commun.\ Math.\ Phys.\ {\bf 162} (1994) 147-174,
hep-th/9303038
%%CITATION = HEP-TH 9303038;%%
%
\bibitem{Chari}
V.~Chari and A.~N.~Pressley,
``A guide to quantum groups,"
Cambridge University Press (1995)
%
\bibitem{Drinfeld2}
V.~G.~Drinfel'd,
``Quantum groups,''
Proc.\ ICM, MSRI, Berkeley (1986) 708
%
\bibitem{Drinfeld1}
V.~G.~Drinfel'd,
``On Poisson homogeneous spaces of Poisson-Lie groups,''
Teoreticheskaya i Mathematicheskaya Fizika, {\bf 95}, 2 (1993) 226-227,
translated in Theoretical and Mathematical Physics, Springer Verlag New York, {\bf 95}, 2 (1993) 524-525
%
\bibitem{Hlavaty1}
L.~Hlavat\'{y} and L.~\v{S}nobl,
``Classification of Poisson-Lie T-dual models with two-dimensional targets,"
Mod.\ Phys.\ Lett.~A {\bf 17} (2002) 429-434,
hep-th/0110139
%%CITATION = HEP-TH 0110139;%%
%
\bibitem{Hlavaty2}
L.~Hlavat\'{y} and L.~\v{S}nobl,
``Classification of 6-dimensional real Manin triples,''
math.QA/0202209
%%CITATION = MATH-QA 0202209;%%
%
\bibitem{Hlavaty3}
L.~Hlavat\'{y} and L.~\v{S}nobl,
``Classification of 6-dimensional real Drinfeld doubles,''
Int.\ J.\ Mod.\ Phys.~A {\bf 17} (2002) 4043-4068,
math.QA/0202210
%%CITATION = MATH-QA 0202210;%%
%
\bibitem{KS2}
C.~Klim\v{c}\'{\i}k and P.~\v{S}evera,
``Poisson-Lie T-duality: open strings and D-branes,''
Phys.\ Lett.~B {\bf 376} (1996) 82-89,
hep-th/9512124
%%CITATION = HEP-TH 9512124;%%
%
\bibitem{Stanciu}
S.~Stanciu,
``D-branes in group manifolds,"
JHEP {\bf 0001} (2000) 025,
hep-th/9909163
%%CITATION = HEP-TH 9909163;%%
%
\bibitem{Klimcik2}
C.~Klim\v{c}\'{i}k,
``Nested T-duality,''
Lett.\ Math.\ Phys.\ {\bf 77} (2006) 99,
hep-th/0505240
%%CITATION = HEP-TH 0505240;%%
%
\bibitem{Sfetsos1}
K.~Sfetsos,
``Poisson-Lie T-duality and supersymmetry,''
Nucl.\ Phys.\ {\bf 56B} (Proc.\ Suppl.) (1997) 302-309,
hep-th/9611199
%%CITATION = HEP-TH 9611199;%%
%
\bibitem{vonUnge}
E.~Tyurin and R.~von~Unge,
``Poisson-Lie T-duality: the path-integral derivation,"
Phys.\ Lett.~B {\bf 382} (1996) 233-240,
hep-th/9512025
%%CITATION = HEP-TH 9512025;%%
%
%
\bibitem{KS4}
C.~Klim\v{c}\'{i}k and P.~\v{S}evera,
``Poisson-Lie T-duality and Loop Groups of Drinfeld Doubles,''
Phys.\ Lett.~B {\bf 372} (1996) 65,
hep-th/9512040
%%CITATION = HEP-TH 9512040;%%
%
\bibitem{ALZ1}
C.~Albertsson, U.~Lindstr\"om and M.~Zabzine,
``N=1 supersymmetric sigma model with boundaries, I,''
Commun.\ Math.\ Phys.\ {\bf 233} (2003) 403,
hep-th/0111161
%%CITATION = HEP-TH 0111161;%%
%
\bibitem{ALZ2}
C.~Albertsson, U.~Lindstr\"om and M.~Zabzine,
``N=1 supersymmetric sigma model with boundaries, II,''
Nucl.\ Phys.\ B {\bf 678} (2004) 295-316,
hep-th/0202069
%%CITATION = HEP-TH 0202069;%%
%
\bibitem{ALZ3}
C.~Albertsson, U.~Lindstr\"om and M.~Zabzine,
``Superconformal boundary conditions for the WZW model,''
JHEP {\bf 0305} (2003) 050,
hep-th/0304013
%%CITATION = HEP-TH 0304013;%%
%
\bibitem{ALZ4}
C.~Albertsson, U.~Lindstr\"om and M.~Zabzine,
``T-duality for the sigma model with boundaries,''
JHEP {\bf 0412} (2004) 056,
hep-th/0410217
%%CITATION = HEP-TH 0410217;%%
%
\bibitem{Sfetsos2}
K.~Sfetsos,
``Canonical equivalence of non-isometric sigma models and
Poisson-Lie T-duality,''
Nucl.\ Phys.~B {\bf 517} (1998) 549-566,
hep-th/9710163
%%CITATION = HEP-TH 9710163;%%
%
\bibitem{Sfetsos3}
K.~Sfetsos,
``Poisson-Lie T-duality beyond the classical level and the renormalization
group,''
Phys.\ Lett.~B {\bf 432} (1998) 365-375,
hep-th/9803019
%%CITATION = HEP-TH 9803019;%%
%
\bibitem{Yano}
K.~Yano and M.~Kon,
``Structures of manifolds,"
Series in Pure Mathematics, Vol.~3,
World Scientific, Singapore (1984)
%
\bibitem{Klimcik9}
C.~Klim\v{c}\'{i}k,
``Poisson-Lie T-duality,''
Nucl.\ Phys.\ Proc.\ Suppl.\ {\bf 46} (1996) 116-121,
hep-th/9509095
%%CITATION = HEP-TH 9509095;%%
%
\bibitem{KS3}
C.~Klim\v{c}\'{i}k and P.~\v{S}evera,
``T-duality and the moment map,''
hep-th/9610198
%%CITATION = HEP-TH 9610198;%%
%
\bibitem{Hassan}
S.~F.~Hassan,
``T-duality and non-local supersymmetries,''
Nucl.\ Phys.\ B {\bf 460} (1996) 362-378,
hep-th/9504148
%%CITATION = HEP-TH 9504148;%%
%
\bibitem{vonUnge2}
R.~von~Unge,
``Poisson-Lie T-plurality,''
JHEP {\bf 0207} (2002) 014,
hep-th/0205245
%%CITATION = HEP-TH 0205245;%%
%
\end{thebibliography}
\end{document}